\documentclass[magazine]{IEEEtran}
\usepackage{fontawesome5}
\usepackage{color}
\usepackage{graphicx, subfigure}
\usepackage{amsmath}
\usepackage{cite}
\usepackage{amsfonts}
\usepackage{multirow,colortbl}
\usepackage{url}
\usepackage{threeparttable}
\usepackage{rotating}
\usepackage{hyperref}
\usepackage{booktabs}
\usepackage{makecell} 
\usepackage[many]{tcolorbox}

\begin{document}
 
\title{Large AI Models for Wireless Physical Layer}
\author{ {Jiajia~Guo, \IEEEmembership{\normalsize {Member,~IEEE}},
Yiming~Cui,
Shi Jin, \IEEEmembership{\normalsize {Fellow,~IEEE}}
and Jun Zhang, \IEEEmembership{\normalsize {Fellow,~IEEE}}
}
\thanks{J. Guo and J. Zhang are with the Department of Electronic and Computer Engineering, The Hong Kong University of Science and Technology, Hong Kong (Email: jiajiaguo@seu.edu.cn, eejzhang@ust.hk).}
\thanks{Y. Cui and S. Jin is with National Mobile Communications Research Laboratory, Southeast University, Nanjing 210096, P. R. China (Email:  cuiyiming@seu.edu.cn, jinshi@seu.edu.cn).}
}

 
\maketitle

\begin{abstract}

Large artificial intelligence models (LAMs) are transforming wireless physical layer technologies through their robust generalization, multitask processing, and multimodal capabilities. This article reviews recent advancements in applying LAMs to physical layer communications, addressing obstacles of conventional AI-based approaches. LAM-based solutions are classified into two strategies: leveraging pre-trained LAMs and developing native LAMs designed specifically for physical layer tasks. The motivations and key frameworks of these approaches are comprehensively examined through multiple use cases. Both strategies significantly improve performance and adaptability across diverse wireless scenarios. Future research directions, including efficient architectures, interpretability, standardized datasets, and collaboration between large and small models, are proposed to advance LAM-based physical layer solutions for next-generation communication systems.
\end{abstract}

\IEEEpeerreviewmaketitle
\vspace{-0.6cm}
 
\section{Introduction}
\label{s1}
Artificial intelligence (AI), especially deep learning, has undergone remarkable advancements over the past decade. The integration of AI with wireless communications has emerged as a transformative paradigm, significantly enhancing capabilities across various network layers, notably the wireless physical layer \cite{cui_overview_2025}.
Unlike the traditional physical layer algorithms designed using rigorous mathematical proofs, AI-based approaches leverage extensive wireless data to train neural networks (NNs), enabling them to directly perform the functions of specific or multiple communication modules effectively, such as channel estimation/prediction, channel state information (CSI) feedback, beamforming, and signal detection
This emerging paradigm has been extensively investigated by academia and industry in recent years and is recognized as a pivotal component in next-generation mobile communications. For example, since 2021, the 3rd Generation Partnership Project (3GPP) has explored AI-based techniques for CSI feedback, beam management, and localization, with plans to standardize these technologies from {day} 1 of 6G.

Compared to computer vision (CV), natural language processing (NLP), and other wireless network layers, the physical layer imposes significantly stricter algorithmic requirements, such as high generalization and robustness. Despite substantial advancements in AI-based wireless physical layer techniques, these approaches face critical challenges, particularly due to their inherent dependence on learning from large volumes of training data.
Specifically, the following obstacles have been recognized in current AI-based wireless physical layer.

\begin{enumerate}
    \item[\textbullet]{\bf Obstacle 1: Complex NN design---}AI for the physical layer eliminates manual communication algorithm design, yet the intensive, task- and scenario-specific manual NN architecture design it requires often impedes deployment.
    \item[\textbullet]{\bf Obstacle 2: Poor generalization---}{Practical deployments often face unavoidable mismatches between training and real-world data distributions arising from differences in scenarios, configurations, etc., degrading NN performance and undermining physical layer reliability.}
    \item[\textbullet]{\bf Obstacle 3: Immense data dependency---}Achieving exceptional physical layer NN performance relies on robust training with extensive, diverse data, yet data collection consumes resources and raises privacy concerns.
    \item[\textbullet]{\bf Obstacle 4: Constrained multimodal processing capabilities---}The {current compact} AI-based physical layer struggles to fully exploit performance gains from other-modal information, particularly sensing data, due to the constrained multimodal processing capabilities.
\end{enumerate}
{Furthermore, while the computational complexity of NNs remains a significant challenge, the advent of powerful accelerating hardware, combined with model compression methods, enables these models to operate with considerable potential.}

In alignment with scaling laws, NN model performance improves with larger models, expanded datasets, and increased computational capacity.
In recent years, large AI models (LAMs), including large language models (LLMs), large vision models (LVMs), and large multi-modal models (LMMs), have excelled in CV and NLP. This success is attributed to their ability to learn powerful representations from extensive, large-scale datasets, enabling the models to capture intricate patterns and relationships within the data \cite{zhu_wireless_2025}.
Inspired by the success of LAMs in domains such as NLP and CV, researchers have applied these models to the wireless physical layer,  addressing the above four challenges within conventional AI-based physical layer approaches.
Existing research in this area can be broadly categorized into two approaches. The first leverages powerful LAMs, pre-trained on language and vision data, to enhance various modules of physical layer, including channel prediction \cite{liu_llm4cp_2024}, beam prediction \cite{sheng_beam_2025}, and CSI feedback \cite{cui_exploring_2025}. The second approach focuses on constructing LAMs specifically for the wireless physical layer from the ground up, utilizing extensive wireless data, such as \cite{guo_prompt2025,alikhani_large_2025,yang_wirelessgpt_2025}.
Both approaches have demonstrated significant potential for advancing wireless physical layer performance.

This article provides a comprehensive review of recent studies on LAMs that focus on the physical layer, covering both the utilization of pre-trained LAMs and the development of native LAMs specifically designed for physical layer. It also seeks to highlight the potential and challenges of physical layer LAMs while providing guidance for future research.
The remainder of this article is structured as follows. The next section provides a concise overview of the fundamental concepts of LAMs {and compares pre‑trained and wireless‑native approaches}. Subsequently, we present several application examples of leveraging pre-trained LAMs, including LLMs and LVMs, to enhance physical layer performance. This is followed by a discussion of physical layer-native LAMs, which are trained from scratch using extensive wireless datasets.
The final section concludes the article by outlining research directions for physical layer LAMs.

{{\emph{Note: }}Given the focused scope of this  paper on LAMs and their advantages over conventional small AI models, we do not provide direct comparisons between AI-based methods and non-AI baselines here. Interested readers are referred to \cite{cui_overview_2025} for detailed performance comparisons in this regard.}

\section{Brief Overview of LAMs}
\label{s2}
In this section, we first introduce the fundamentals of LAMs and then highlight their key advantages, which hold significant potential for advancing their application in the wireless physical layer.
\subsection{Basic Ideas of LAMs}

LAMs rely on three core elements: computing power, algorithms, and data. These models are constructed from numerous stacked NNs, featuring billions or even trillions of parameters and intricate computational architectures. This design significantly enhances their expressive and predictive capabilities, enabling them to process vast datasets and tackle highly complex tasks.
Training on extensive datasets empowers LAMs with the capacity for effective generalization, as they autonomously discern intricate patterns and features from raw data. This remarkable ability to independently extract high-level patterns, which often appears unexpectedly as models scale, is a phenomenon termed ``emergence.'' 

While first conceived for NLP following the introduction of the Transformer architecture in 2017, these models—epitomized by LLMs like ChatGPT—have undergone a significant evolution. They have since emerged as a foundational paradigm across the broader field of AI, their design enabling broad applicability from areas such as CV to complex multi-modal tasks. For instance, GPT-4o, released in 2024, exemplifies this multi-modal capability by accepting any combination of text, audio, image, and video inputs to generate diverse outputs across text, audio, and image modalities.

\subsection{Key Advantages of LAMs}

Owing to extensive computational resources, advanced algorithms, and large-scale datasets, LAMs significantly outperform smaller AI models, and their primary advantages {over current smaller models} are outlined below.
\begin{enumerate}
    \item[\textbullet]{\bf Advantage 1: Powerful generalization capabilities---}Through training on immense datasets, LAMs build a deep, versatile understanding of patterns and relationships within diverse information, enabling them to effectively process and interpret previously unseen data across various scenarios.

      \item[\textbullet]{\bf Advantage 2: Robust multi-task processing ability---}The vast number of NN parameters provides LAMs with the powerful capacity needed to effectively from massive amounts of data, leading to their impressive performance across a wide range of AI tasks.

      \item[\textbullet]{\bf Advantage 3: Excellent few-shot learning---}Extensive training on diverse data help LAMs build a deep understanding of underlying patterns, thereby significantly reducing their reliance on large amounts of new data. Therefore, pre-trained LAMs can quickly adapt to new tasks, performing well with few examples.

       \item[\textbullet]{\bf Advantage 4: Superior multimodal processing ability---}Through the inherent capacity of LAMs' NNs for high-dimensional representations, massive scaling induces emergent cross-modal grounding, which allows patterns from distinct modalities to interact and translate within a unified processing framework.
        \item[\textbullet]{\bf Advantage 5: Hardware friendliness---}LAMs employ highly parallel and regularly structured  NN architectures like Transformers. Their core operations (like massive matrix computations) perfectly align with the design of specialized AI hardware, leading to exceptional computational efficiency and speed.
\end{enumerate}

These inherent advantages of LAMs over smaller AI models offer a promising solution to the fundamental obstacles of conventional AI-physical layer approaches, thus enabling the construction of high-performance, strongly generalized flexible physical layers for real-world deployment.

\begin{table*}[!t]
\caption{Comprehensive Summary of Pre-trained LAM Applications in Wireless Physical Layer}
\label{table_summary}
\scriptsize
\centering
\resizebox{\textwidth}{!}{
\begin{tabular}{c|c|c|c|c|c|l|c}
\toprule
\textbf{Method} & \textbf{PHY Task} & \makecell[c]{\textbf{Pre-trained}\\ \textbf{LAM}} & \textbf{Preprocessing} & \makecell[c]{\textbf{Output}\\ \textbf{layers}}& \textbf{Fine-tuning} & \makecell[c]{\textbf{Key}\\ \textbf{Contributions}}  & \textbf{Date} \\
\hline \hline
LLM4CP \cite{liu_llm4cp_2024} & Channel pred. & GPT-2 & Reshape/patching, embed & \checkmark & \checkmark & First applying pre-trained LLMs to PHY layer & 06/2024 \\
\hline
BP-LLM \cite{sheng_beam_2025} & Beam pred. & GPT-2 & Instance norm/patching, embed & \checkmark &   $\times$ & Introducing Prompt-as-Prefix mechanism & 08/2024 \\
\hline
\makecell[c]{Multi-task\\LLM \cite{zheng_large_2025}} & \makecell[c]{Precoding,\\signal detection,\\channel pred.} & LLAMA2 & \makecell[c]{LLM embedder,\\wireless embedding} & \checkmark & \checkmark & One pre-trained LLM for multile tasks & 12/2024 \\
\hline
LLM \cite{cui_exploring_2025} & CSI feedback & GPT-2 & Reshape/patching, embed& \checkmark & \checkmark & Introducing LLM-based language polish for CSI & 01/2025 \\
\hline
2DLAM \cite{xie_2dlam_nodate}& \makecell[c]{Delay-Doppler \\estimation} & ImageGPT & Feature extract, embed& \checkmark & $\times$ & Introducing LVMs to PHY layer & 03/2025 \\
\hline
LLM4SG \cite{han_llm4sg_2025} & \makecell[c]{Scatterer\\ generation} & GPT-2 & \makecell[c]{LiDAR process,\\ feature extract, embed} & \checkmark & \checkmark & Leveraging LLMs for LiDAR-to-scatterer mapping & 06/2025 \\
\hline
M$^2$BeamLLM \cite{zheng_m2beamllm_2025}& Beam pred. & GPT-2 & \makecell[c]{Data encoding,\\  multimodal data fusion} & \checkmark & \checkmark & Multimodal integration & 06/2025 \\
\hline
LLM \cite{xu_llm-empowered_2025}& \makecell[c]{Far/near-field\\class., precoding} & GPT-2 & \makecell[c]{CSI rearrange/norm,\\ feature extract, embed} & \checkmark & \checkmark & Regarded LLMs as a general wireless solver & 06/2025 \\
\hline
LVM4CSI \cite{guo_lvm4csi_2025}& \makecell[c]{Channel est.,\\ HAR, localization} & \makecell[c]{ConvNeXt,\\ DINO-X} & CSI-to-CV transformation& \checkmark/$\times$ & $\times$ &Proposing LVM-enabled framework for CSI tasks & 07/2025 \\
\bottomrule
\end{tabular}}
\\
\vspace{2mm}
\footnotesize
\textbf{Note:} pred.: prediction; embed: embedding; norm: normalization; gen.: generation; est.: estimation; HAR: Human activity recognition.
\vspace{-6mm}
\end{table*}

{
\subsection{Comparative Framework: Pre‑trained vs. Wireless-Native}
To orient readers before Sections \ref{s3} and \ref{s4}, this subsection provides a concise framework for choosing between pre-trained LAMs and wireless‑native LAMs. Pre‑trained LAMs adapt general‑purpose models from NLP/CV to physical‑layer tasks via data alignment and task-specific fine‑tuning. Wireless‑native LAMs are trained on wireless signals and scenarios to learn domain-consistent representations.

Selecting between pre-trained and wireless‑native LAMs primarily depends on two factors: modality alignment and computational complexity with real‑time viability. When wireless data such as CSI can be aligned to the source domains of pre‑trained models and latency constraints are relaxed, pre‑trained LAMs enable rapid prototyping and effective multimodal integration with limited labeled data. Under stringent physical-layer latency constraints and device-side power and energy budgets, wireless‑native LAMs, designed with explicit complexity constraints together with model compression and hardware-aware optimization, are preferred. This framing guides the detailed discussions in Sections \ref{s3} and \ref{s4}.
For practitioner assessment of the computational complexity of wireless‑native LAMs, see Section \ref{4d} (Table \ref{tab:performance}); the complexity of pre‑trained backbones is mentioned in Section \ref{3d}.
The following sections elaborate on these two strategies with representative frameworks and use cases.
}

\section{Pre-trained LAMs for Physical Layer}
\label{s3}

\subsection{Motivation}

Owing to the powerful generalization capabilities, excellent few-shot learning and powerful multimodal processing ability of LAMs (\textbf{Advantages 1,  3\&4}), pre-trained LAMs can be effectively transferred to new scenarios or (multimodal) tasks through fine-tuning with limited training samples. For instance, while LLMs like DeepSeek-R1 cannot directly access sensitive banking records due to data security protocols, they can be locally fine-tuned with minimal transaction data to accurately detect suspicious activity.

These advantages offer a promising solution to address the challenges of poor generalization immense data dependency, and constrained multimodal processing
capabilities {in conventional/small-scale AI approaches} (\textbf{Obstacles 2, 3 \&4}). In conventional AI-based physical layer approaches, CSI and other wireless data are often treated as ``image'' or sequence data, processed by NNs originally developed for CV and NLP. Indeed, these wireless data exhibit high structural similarity to data in CV and NLP. The extensive knowledge acquired by LAMs through vast training datasets can therefore be highly beneficial for physical layer tasks. By inheriting this knowledge, physical layer NNs may require significantly less training data and achieve substantially higher generalization capabilities.

\begin{figure}[t]
    \centering    \includegraphics[width=0.5\textwidth]{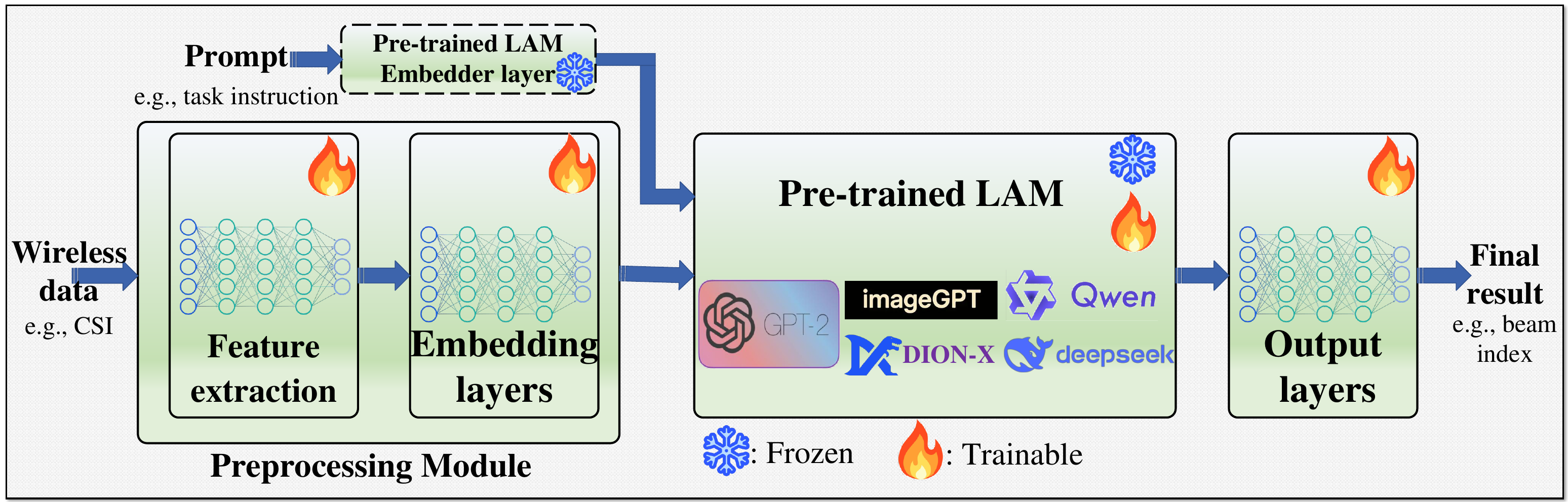}
    \caption{\label{LLM4CP}General framework of pre-trained LAMs for physical layer. The framework adapts LAMs using a preprocessing module to align wireless data with input space and output layers to match physical-layer tasks. Depending on the task, LAM parameters are selectively fine-tuned \cite{liu_llm4cp_2024} or frozen \cite{guo_lvm4csi_2025}.}
    \vspace{-6mm}
\end{figure}

\subsection{Key Framework}

Fig. \ref{LLM4CP} illustrates a general framework of pre-trained LAMs for wireless physical layer. This framework, initially proposed by \cite{liu_llm4cp_2024}, has been widely adopted by most existing works.
This framework consists of three main components, as follows.

{\bf Preprocessing module: }Wireless data, distinct from language and CV modalities, cannot be directly input into pre-trained LAMs and requires preprocessing to align with the LAM's input format. As most LAMs use a Transformer backbone processing tokenized inputs, the preprocessing module (left part of Fig. \ref{LLM4CP}) includes two steps: feature extraction and embedding. Feature extraction uses tailored NN architectures, like stacked FC or convolutional layers in \cite{xie_2dlam_nodate}, to capture wireless data characteristics. Learnable embedding layers, with attention mechanisms, FC layers, and positional encoding, transform these features into tokens for Transformer-based LAM processing.

{
{\bf Pre-trained LAM: }Pre-trained LAMs, trained on extensive datasets, acquire broad general knowledge that provides universal modeling capabilities. These models can serve as backbone NN modules for various physical-layer tasks.
However, because of the substantial gap between the data used to train LAMs and wireless-specific datasets, directly applying pre-trained LAMs for feature extraction is often suboptimal. In most cases, fine-tuning is necessary to adapt the models and improve performance. Trainable components such as attention layers are commonly adjusted to capture wireless-specific characteristics, as shown in LLM-based approaches \cite{liu_llm4cp_2024}.
Freezing pre-trained LVMs is considered only when the domain gap is small, for example when CSI is treated as images in LVMs \cite{guo_lvm4csi_2025}. This approach helps further preserve generalization (Advantage 1), enables few-shot efficiency (Advantage 3), and reduces computational cost, but it remains secondary compared with fine-tuning in most scenarios.}

{\bf Output layers: }The output module is designed to convert the output features of pre-trained LAMs into final prediction results customized for specific physical layer task types.
For instance, in the beam prediction task, which can be treated as a standard classification task, the output layer typically incorporates multiple FC layers followed by a softmax layer \cite{sheng_beam_2025}. Conversely, for channel denoising or prediction tasks, only FC layers are required to align the LAM's output features with high-quality CSI.

The training of the physical layer NNs, powered by pre-trained LAMs, adopt an end-to-end approach. The preprocessing and output modules, being entirely new NNs without pre-trained parameters, must be trained from scratch. For pre-trained LAMs, as previously noted, their parameters can remain frozen throughout training if the wireless data closely resembles the LAMs' training data. Otherwise, the LAMs' parameters require fine-tuning alongside the preprocessing and output NN layers.

\subsection{Use Cases}
Since June 2024 \cite{liu_llm4cp_2024}, several studies have explored the application of pre-trained LAMs to the wireless physical layer. As presented in Table \ref{table_summary}, these studies encompass pre-trained LAMs for tasks such as channel prediction, beam prediction, precoding, signal detection, and CSI feedback. The significant potential of pre-trained LAMs in the physical layer is beginning to emerge. In this section, we present three representative use cases to demonstrate their functionality, including pre-trained LLMs for channel prediction \cite{liu_llm4cp_2024}, pre-trained LVMs for CSI tasks \cite{guo_lvm4csi_2025}, and pre-trained LLMs for multiple physical layer tasks \cite{zheng_large_2025}.

\begin{figure}[t]
    \centering    \includegraphics[width=0.5\textwidth]{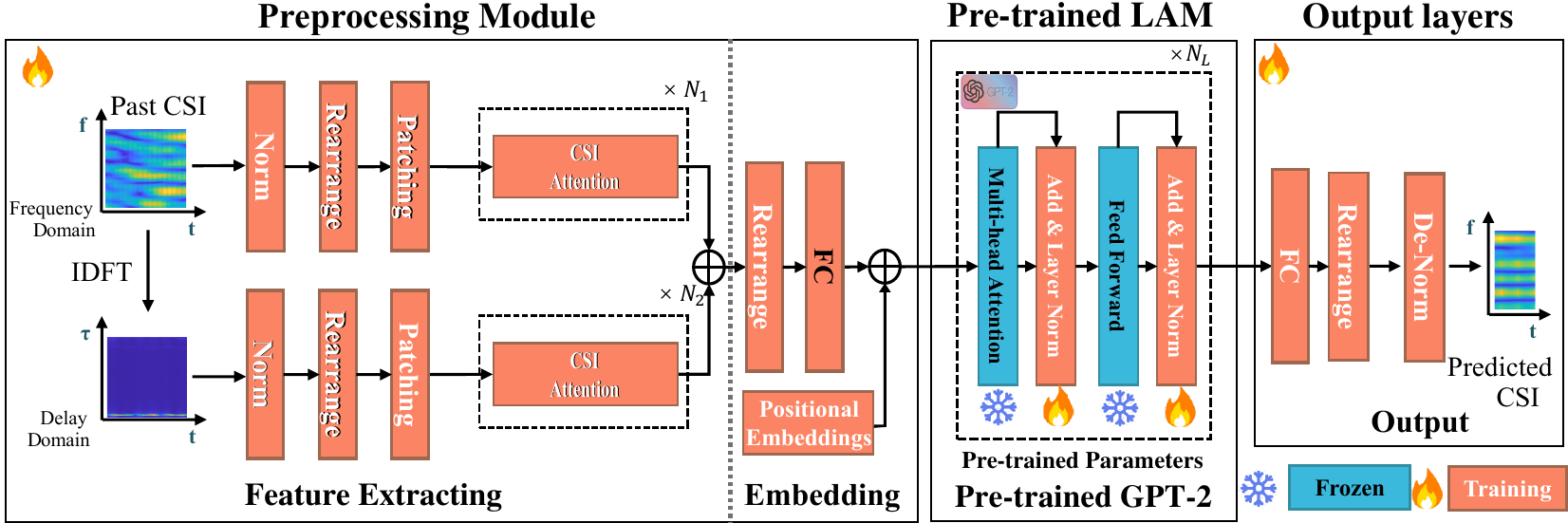}    \caption{\label{LLM4CPnn}{Detailed architecture of the LLM4CP framework proposed in \cite{liu_llm4cp_2024}.}}
    \vspace{-6mm}
\end{figure}
\subsubsection{Pre-trained LLMs for Channel Prediction}AI-based channel prediction significantly reduces CSI acquisition overhead in channel estimation and feedback, with studies ongoing in 3GPP since 2021 \cite{cui_overview_2025}. This approach predicts future CSI using historical data across the same or distinct frequency bands. However, in high-velocity scenarios with bidirectional links on different bands, existing AI-based methods struggle to meet accuracy requirements due to high problem complexity and limited NN capabilities and often exhibit poor generalization when scenarios change. To address these issues, \cite{liu_llm4cp_2024} proposes LLM4CP, a novel framework leveraging the expressive power of pre-trained LLMs.

The main framework of LLM4CP is illustrated in Fig. \ref{LLM4CPnn}.
To reduce the NN complexity, the LLM4CP is designed to predict future downlink CSI for each antenna pair using its corresponding uplink CSI sequence, rather than the entire CSI.
Preprocessing involves normalizing, arranging, and patching frequency- and delay-domain CSI sequences with attention blocks, followed by an NN-based embedding module to generate tokens. These tokens are processed by a pre-trained GPT-2 model, with an FC layer and denormalization to align outputs with the future CSI format. During training, to enhance training efficiency and reducing costs, only specific GPT-2 components, including addition, layer normalization, and positional embedding, are fine-tuned to adapt the LLM for channel prediction, while multi-head attention and feed-forward layers remain frozen to preserve universal knowledge. Across full-sample, few-shot, and generalization tests, simulations confirm LLM4CP's superior performance against non-AI and other AI methods, at acceptable training and inference costs.

\begin{figure}[t]
    \centering
    \includegraphics[width=0.45\textwidth]{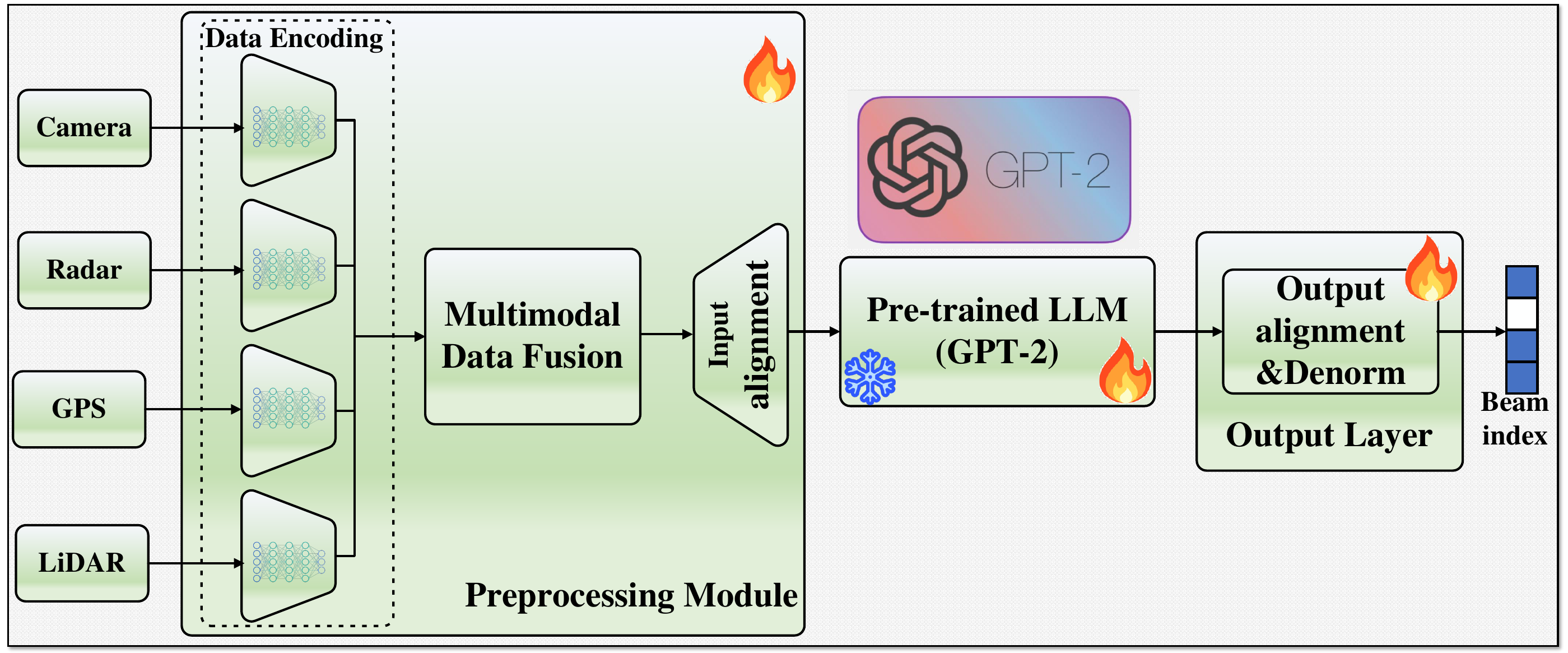}
    \caption{\label{M2BeamLLM}Main framework of the pre-trained LLMs for multimodal sensing-empowered beam prediction, i.e., M$^2$BeamLLM \cite{zheng_m2beamllm_2025}, comprising three modules, akin to the general framework illustrated in Fig. \ref{LLM4CP}. }
    \vspace{-6mm}
\end{figure}
\subsubsection{Pre-trained LVMs for CSI Tasks}CSI is fundamental to the wireless physical layer, with its acquisition and utilization being critical components. Pre-trained LLMs, such as GPT-2, have been employed to improve CSI-related task performance and generalization, leveraging their robust adaptability across diverse tasks \cite{liu_llm4cp_2024,  cui_exploring_2025}. However, the modality gap between CSI and language data necessitates time- and computation-intensive fine-tuning. By visualizing CSI samples in the angular-delay domain, the authors in \cite{guo_lvm4csi_2025} identifies strong structural similarities with images, proposing LVM4CSI---a pre-trained LVM-based framework that eliminates fine-tuning for CSI-related tasks. 

Unlike pre-trained LLM-based approaches that rely on NNs to generate tokens for LLMs, the preprocessing module of LVM4CSI, proposed in \cite{guo_lvm4csi_2025}, transforms CSI data into a CV-compatible image format. For RGB image representation, operations like MATLAB’s \emph{jet} colormap can directly colorize grayscale CSI samples when only magnitude information is required. Alternatively, a blank channel can be added to a two-channel CSI sample, with real and imaginary parts as separate channels.
The RGB CSI ``image'' is then processed by LVMs (e.g. ConvNeXt) to extract features, followed by a streamlined output module to align these features with the target outcomes.
During training, only the NNs in the output module are trained, significantly reducing computational overhead. Simulation results demonstrate that the LVM4CSI achieves comparable or superior performance to task-specific NNs.

\subsubsection{Pre-trained LLMs for Multimodal Sensing-empowered Beam Prediction}
Beam prediction depends on the propagation environment between the base station and user, captured by sensors such as cameras, radar, GPS, and LiDAR. Beam prediction performance can be significantly improved by utilizing data from as many sensors as possible.
However, the diverse modalities of these sensor data present significant challenges in efficiently integrating them simultaneously.
Pre-trained LAMs, leveraging extensive multimodal data, exhibit strong potential for processing diverse data inputs. Accordingly, to enhance beam prediction performance, the authors in \cite{zheng_m2beamllm_2025} propose M$^2$BeamLLM, a multimodal sensing-empowered beam prediction framework based on pre-trained LLMs.

Fig. \ref{M2BeamLLM} depicts the M$^2$BeamLLM framework proposed in \cite{zheng_m2beamllm_2025}, comprising three modules, akin to the general framework illustrated in Fig. \ref{LLM4CP}. The preprocessing module first encodes sensor data from diverse sources, then fuses and aligns them with the LLM’s input. Similar to LLM4CP, the aligned input is processed by a pre-trained GPT-2 model, with select components frozen and others fine-tuned. The LLM’s output is finally mapped to the beam index. Simulations demonstrate that incorporating sensor data enhances beam prediction performance, with pre-trained LLMs significantly improving accuracy and robustness.

{
\subsection{Quantitative Complexity Analysis}
\label{3d}
Pre-trained LAMs (e.g., GPT, LLaMA2, ImageGPT) employed as backbones for wireless feature extraction generally incur substantially higher computational complexity than compact baselines. For example, 2DLAM~\cite{xie_2dlam_nodate} built on ImageGPT contains 172.56 M parameters \cite[Table~I]{xie_2dlam_nodate}, implying significantly greater per-inference computation and latency. Consequently, such models are typically unsuitable for stringent real-time physical-layer tasks, while aligning better with relaxed-latency sensing or offline pipelines.
}

\begin{figure}[t]
    \centering
    \includegraphics[width=0.5\textwidth]{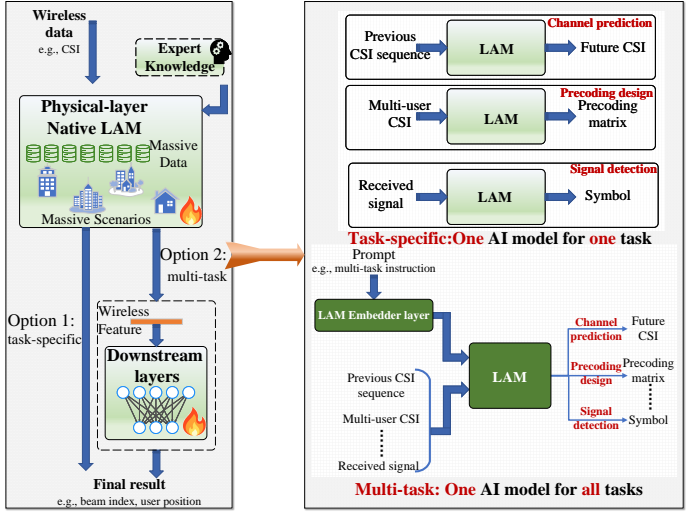}
    \caption{\label{NativeFramework}{A general framework of wireless physical layer-native LAMs. This framework trains an LAM from scratch for a specific \cite{guo_prompt2025, catak_bert4mimo_2025} or multiple physical tasks \cite{alikhani_large_2025,yang_wirelessgpt_2025}.}}
    \vspace{-6mm}
\end{figure}
 
\section{Wireless Physical Layer-native LAMs}
\label{s4}
\subsection{Motivation} 
While adapting pre-trained LAMs from domains like CV and NLP offers a pragmatic approach to leverage their powerful capabilities, a fundamental challenge remains: the inherent modality gap between general data (e.g., natural images and text) and wireless signals. Wireless data, such as CSI, possesses unique physical properties (e.g., complex-valued numbers, spatial-temporal-frequency correlations, and sparsity in specific domains) that are not present in the datasets used to train LAMs. This domain gap renders the learned representations from pre-trained LAMs suboptimal for wireless tasks, as they may not fully capture the underlying wireless physics. 
{Moreover, as existing LLMs and LVMs are designed to address a wide range of tasks with a single model, their complexity is excessive for wireless physical layer applications, rendering them impractical for real-time deployment.}

To overcome these limitations, a more direct and potentially more powerful approach has emerged: building native LAMs for the wireless physical layer from the ground up. 
This approach enables the model to generate universal, rich, and contextualized feature representations (embeddings) of the wireless environment for the utilization of downstream tasks with minimal adaptation. This aligns perfectly with the LAM's key advantages, particularly robust multi-task processing and excellent few-shot learning ({\bf Advantages 2 \& 3}). Moreover, as LAMs rely minimally on specific NN architectures and incorporate widely adopted Transformer blocks, they can effectively address the challenges of complex NN designs ({\bf Obstacle 1}).

\subsection{Key Framework}
Fig. \ref{NativeFramework} depicts a general framework for physical-layer native LAMs. Their development mirrors that of LAMs in NLP and CV, relying on extensive data, robust NNs, and corresponding training as foundational elements.

{\bf Extensive data: }NNs derive their knowledge from training data. Existing AI-based approaches collect or generate wireless data from specific regions, training NNs solely with these samples, which limits their applicability to other environments. To develop a robust LAM for the wireless physical layer, data must be gathered from diverse scenarios. However, the superior performance of certain tasks, such as CSI feedback, relies on extracting and leveraging environment-specific features. Thus, to ensure optimal performance, this factor must be addressed. Integrating expert knowledge (scenario knowledge) may provide a solution to this challenge.

{\bf Robust NNs: }Numerous novel NN architectures have been proposed for physical layer tasks. However, research in NLP and CV demonstrates that, with extensive training data, stacked Transformer blocks outperform other meticulously designed NN architectures. Consequently, to develop a robust physical layer LAM, most related studies directly adopt Transformer blocks, offering a simple yet effective approach.

{\bf Task-specific and universal LAMs: }
As depicted in Fig. \ref{NativeFramework}, physical-layer native LAMs can be designed for specific tasks, including channel prediction, precoding design, and signal detection. Also, leveraging the robust capabilities of LAMs, a single model can effectively perform multiple tasks \cite{zheng_large_2025} or extract features for multiple tasks \cite{alikhani_large_2025,yang_wirelessgpt_2025}.

 \begin{table}[t]
\centering
\caption{\label{tab:performance}{Performance and complexity comparison of different CSI feedback AI models \cite{guo_prompt2025}}}
\renewcommand{\arraystretch}{1.2}
\begin{tabular}{lcccc}
\toprule
\multirow{2}{*}{\textbf{Metric}} & \multicolumn{4}{c}{\textbf{Methods}} \\
\cmidrule(lr){2-5}
 & \textbf{Tiny} & \textbf{Simple} & \textbf{LAM} & \textbf{Prompt+LAM} \\
\midrule
\rowcolor[gray]{0.95}
NMSE (dB) & -3.05 & -4.59 & -6.71 & -8.36 \\
Runtime CPU (ms) & 3.36 & 5.20 & 20.53 & 21.57 \\
\rowcolor[gray]{0.95}
Runtime GPU (ms) & 0.90 & 1.14 & 2.55 & 2.64 \\
Parameter number & 357,476 & 563,300 & 2,546,020 & 6,705,508 \\
\rowcolor[gray]{0.95}
FLOPs (M) & 29.17 & 57.77 & 334.24 & 342.56 \\
\bottomrule
\end{tabular}
\vspace{-4mm}
\end{table}

\subsection{Use Cases}
The development of native LAMs follows two distinct strategic paths with different frameworks. The first focuses on task-specific models for performance optimization, while the second aims to build universal models for broad applicability. The following use cases illustrate these two approaches.

\subsubsection{Task-specific LAM for Wireless Physical Layer}
AI-based CSI feedback with an autoencoder framework, studied by 3GPP since 2021, has shown significant potential in reducing feedback overhead and enhancing feedback accuracy. The conventional AI-based feedback NN is trained with CSI data from a certain scenario and evaluated with the data with a similar distribution. This causes a serious generalization problem. Inspired by the powerful generalization ability of LAMs and the {mechanism} of AI-based CSI feedback, the authors in \cite{guo_prompt2025} propose a prompt-enabled LAM for CSI feedback, leveraging the advantages of both the LAMs and expert knowledge.

The authors in \cite{guo_prompt2025} examine the factors contributing to efficient CSI feedback through performance analysis on varied wireless datasets. Their findings indicate that exceptional feedback accuracy derives from AI models’ adeptness at data fitting and their capacity to exploit scenario-specific environmental insights. Leveraging these observations, they propose a pioneering LAM for CSI feedback, incorporating a prompt-based mechanism. This LAM utilizes Transformer architectures and is trained on expansive, multi-scenario datasets. Notably, it embeds expert knowledge—specifically, the average channel magnitude in the angular-delay domain—as a decoder prompt to optimize reconstruction quality. 
{At a 144-bit feedback overhead, Table \ref{tab:performance} benchmarks CSI feedback for a prompt‑enabled LAM, its prompt‑free counterpart, and conventional small models. LAMs markedly outperform the small models, with prompts yielding further accuracy gains.}

\subsubsection{Universal LAM for Wireless Physical Layer}

Most methods for tasks in wireless physical layer features supervised learning fashion, which heavily rely on labeled wireless data to train the AI model. However, the scarcity of the labeled data usually limited the practical implementation. Instead of ``teaching'' the AI model with the ``correct answers'', another solution is to employ a powerful LAM to read the raw data to explore the complicated features inside the wireless data without labels. This method is typically implemented with self-supervised training on unlabeled channel data, where a common technique is masked signal modeling. To be specific, the model learns to reconstruct masked portions of the input data, thereby capturing its underlying structure. By using the well-extracted features explored by the LAM, the physical layer tasks can be completed via few additional task-specific layers without introducing significant training overhead.

The divergent philosophies of native LAMs are best illustrated by recent pioneering works. The universal foundation model approach is well-represented by the large wireless model (LWM) \cite{alikhani_large_2025}. LWM is a Transformer-based encoder pre-trained on over a million channel samples using a self-supervised, BERT-like objective. Its primary purpose is to serve as a universal feature extractor. Once pre-trained, its powerful channel embeddings can be used to significantly enhance the performance and data efficiency of various downstream tasks, including beam prediction and LoS/NLoS classification. The work on WirelessGPT follows a similar vision, aiming to build a single foundation model for both communication and sensing applications \cite{yang_wirelessgpt_2025}. This line of research embodies a ``pre-train once, adapt for many'' paradigm.

{
\subsection{Quantitative Complexity Analysis}
\label{4d}
The complexity of wireless LAMs is a primary deployment concern. To provide a concrete view, we analyze the CSI feedback LAM~\cite{guo_prompt2025} as an exemplar. Table~\ref{tab:performance} summarizes the complexity across Transformer-based models that differ in the number of blocks. 
We quantify complexity through \emph{per-inference} CPU/GPU latency, NN parameter numbers, and floating-point operations (FLOPs).

From Table~\ref{tab:performance}, increasing the number of Transformer blocks in LAMs substantially raises both parameters and FLOPs. For example, relative to the tiny model, the LAM's FLOPs are approximately $11.46\times$ higher, and the CPU runtime increases from $3.36$\,ms to $20.53$\,ms. This trend does not persist on the GPU: leveraging parallelism, the LAM's runtime rises only from $0.90$\,ms to $2.55$\,ms, underscoring the efficiency of GPU inference for large-scale NNs. In this case, the modest GPU-side latency increase accompanies a $43\%$ reduction in feedback error compared with the tiny model, which represents a pragmatic trade-off for deployment.}

\section{Future Works and Open Issues} 

Although it has already demonstrated significant promise, the application of LAMs to the wireless physical layer remains in its early stages. To fully unlock this paradigm’s capabilities and enable its practical adoption in 6G and beyond, key challenges and research opportunities must be addressed. Promising directions and open issues for future investigation are outlined below.

\subsection{Efficient, Real-Time, and Hardware-Aware Wireless LAMs}
{
Both pretrained LAMs and wireless-native LAMs face tight latency, power, and memory budgets at the physical layer, especially on the user side where tasks such as CSI processing, beam prediction, and detection often require submillisecond inference under limited energy.
Future work should focus on three concrete fronts: compress and right‑size models; engineer low‑latency inference with efficient dataflow; and align algorithms with accelerators under actual device limits. First, the model size should be reduced through compression and compact designs, using practical distillation, pruning, quantization, and sparsity to reduce parameters and FLOPs while preserving accuracy across diverse scenarios. Second, inference should be accelerated by simplifying core components, organizing dataflow to minimize memory movement, and applying lightweight attention, caching, or early exit where appropriate so execution aligns with physical layer timing. Third, algorithms should be co-designed with accelerators so that precision, operators, and scheduling match real device constraints on platforms like ASIC.}

\subsection{Interpretability of Wireless LAMs} 
For LAMs to be trusted and deployed in mission-critical communication networks, their decision-making processes cannot remain opaque "black boxes." Understanding \textit{why} a model makes a particular prediction is fundamental for debugging, ensuring robustness, and certifying reliability. Consequently, developing eXplainable AI (XAI) techniques specifically for wireless LAMs is a paramount future task. Future work should focus on adapting and creating interpretability methods that are meaningful in the context of wireless communications. For instance, {visualizing the attention mechanisms} within Transformer-based LAMs can reveal which parts of the CSI data the model deems most important for a given task. This could provide valuable physical insights into the model's learned strategies. Furthermore, techniques that assess feature importance could be used to quantify the influence of different channel characteristics on the model's output. A more profound goal would be to probe the internal representations of the LAM to see if they correspond to known physical phenomena (e.g., multipath clusters, angle of arrival/departure). Success in this area would not only build confidence in the models but could also lead to new, AI-driven discoveries about the nature of wireless channels. 

\subsection{Standardized Large-Scale Wireless Datasets} 
The progress of LAMs in NLP and CV has been driven by massive, standardized datasets such as Common Crawl and ImageNet. The wireless community lacks an equivalent resource, hindering reproducibility and fair comparison. Building a truly large-scale, comprehensive, standardized wireless dataset is therefore a priority, and it must exceed the scale of existing efforts.
This dataset should offer billions of channel samples across dense urban, indoor, rural, and high-speed scenarios, span Sub‑6GHz to upper 6 GHz and mmWave, and cover antenna configurations from massive MIMO to extremely large-scale arrays. It should combine high-fidelity outputs from state-of-the-art simulators (for example, ray tracing) with measurements from large real-world campaigns, and be enriched with metadata such as precise location, environmental maps (such as 3D building models), and, where available, LiDAR and camera data. Such a resource would provide a standard benchmark for training and evaluating wireless LAMs and catalyze multimodal research toward a holistic understanding of the communication environment.

\subsection{Large and Small AI Model Collaboration}
LAMs, despite their strong generalization capabilities, cannot fully replace small AI models in wireless physical layer applications. Small models offer distinct advantages, including lightweight architectures and rapid training, enabling efficient adaptation to specific scenarios. To optimize performance, future research should focus on effective collaboration between large and small models, leveraging their complementary strengths. For instance, LAMs can provide robust generalization across diverse environments, while small models excel at quickly learning scenario-specific features, such as those required for CSI feedback or beam prediction. This synergistic approach can enhance accuracy and computational efficiency, tailored to the unique demands of the physical layer. By integrating these models strategically, this paradigm promises to advance 6G systems, balancing scalability with precision in dynamic wireless environments.

\section{Conclusion}

This paper reviews the application of LAMs to the wireless physical layer, highlighting their robust generalization, multitask processing, effective few-shot learning, and advanced multimodal capabilities {that overcome limitations in conventional AI methods}. These attributes enable LAM-based approaches to achieve competitive performance with superior generalization in future communication systems.
Despite this potential, the integration of LAMs into the physical layer remains an emerging field, requiring further exploration. Comprehensive theoretical analyses, practical deployment strategies, and innovative LAM architectures are essential to realize this paradigm in real-world communication scenarios.

\bibliographystyle{IEEEtran}
\bibliography{SelectedLAMpaper}
 \vspace{-6mm}

\begin{IEEEbiographynophoto}
	{Jiajia Guo}[Member, IEEE] (jiajiaguo@seu.edu.cn) received the Ph.D. degree from Southeast University in 2023. His research interests focus on AI-native air interfaces, massive MIMO, ISAC, and large AI models.
\end{IEEEbiographynophoto}
\par\vspace{-0.8cm}
\begin{IEEEbiographynophoto}
	{Yiming Cui}[Graduate Student Member, IEEE] (cuiyiming@seu.edu.cn) is currently working towards his Ph.D. degree with Southeast University.
\end{IEEEbiographynophoto}
\par\vspace{-0.8cm}
\begin{IEEEbiographynophoto}
{Shi Jin}
[Fellow, IEEE] (jinshi@seu.edu.cn) received his Ph.D. degree in information and communications engineering from Southeast University in 2007. He is currently with Southeast University. His research interests include wireless communications, random matrix theory, and information theory.
\end{IEEEbiographynophoto}
\par\vspace{-0.8cm}
 \begin{IEEEbiographynophoto}
	{Jun Zhang}[Fellow, IEEE] (eejzhang@ust.hk) received his Ph.D. degree from the University of Texas at Austin. He is a Professor at HKUST. His research interests include wireless communications and networking, mobile edge computing, and edge AI, and integrated AI and communications
\end{IEEEbiographynophoto}

\end{document}